\documentclass[preprint,12pt]{elsarticle} 
\usepackage{graphicx} 
\usepackage{amssymb} 
\usepackage{amsmath}

\journal{Nuclear Physics A} 

\begin{document} 

\begin{frontmatter} 

%\title{Self-consistent calculations of $\eta$ nuclear bound states} 
\title{In-medium $\eta N$ interactions and $\eta$ nuclear bound states} 
\author[a]{A.~Ciepl\'{y}} 
\author[b]{E.~Friedman} 
%\author[b]{A.~Gal\corref{cor1}} 
%\cortext[cor1]{Corresponding author: avragal@vms.huji.ac.il} 
\author[b]{A.~Gal} 
\author[a]{J.~Mare\v{s}} 
\address[a]{Nuclear Physics Institute, 25068 \v{R}e\v{z}, Czech Republic} 
\address[b]{Racah Institute of Physics, The Hebrew University, 91904 
Jerusalem, Israel} 

\begin{abstract} 

The in-medium $\eta N$ interaction near and below threshold is constructed 
from a free-space chirally-inspired meson-baryon coupled-channel model that 
captures the physics of the $N^{\ast}(1535)$ baryon resonance. Nucleon Pauli 
blocking and hadron self-energies are accounted for. The resulting energy 
dependent in-medium interaction is used in self-consistent dynamical 
calculations of $\eta$ nuclear bound states. Narrow states of width 
$\Gamma_{\eta} \lesssim 2$~MeV are found across the periodic table, 
beginning with $A \geq 10$, for this in-medium coupled-channel interaction 
model. The binding energy of the $1s_{\eta}$ state increases with $A$, 
reaching a value of $B_{1s}(\eta)\approx 15$~MeV. The implications of our 
self-consistency procedure are discussed with respect to procedures used 
in other works.   

\end{abstract}


\end{frontmatter} 

\section{Introduction}
\label{sec:intro}

Haider and Liu realized in 1986 that a moderately attractive $\eta N$ 
interaction, with scattering length estimated as $a_{\eta N}=0.27+{\rm i}
0.22$ fm, may lead to a robust pattern of $\eta$-nuclear bound states across 
the periodic table beginning with $^{12}$C \cite{HL86}. Their pioneering 
work has been followed by numerous studies of the $\eta N$ interaction 
within various theoretical models that yielded a wide range of values 
for Re~$a_{\eta N}$ from 0.2 fm \cite{KWW97} to about 1.0 fm \cite{GW05}, 
as summarized in 2005 by Arndt et al. \cite{Arndt05}. 
Among the very recent works demonstrating this large variation we mention 
the $\pi N$--$\eta N$--$K\Lambda$--$K\Sigma$ coupled-channel chiral model of 
Mai, Bruns, Mei{\ss}ner \cite{MBM12} with values of Re~$a_{\eta N}=0.22$~fm 
and 0.38~fm in its two versions, and the $K$-matrix analysis involving 
additional channels by Shklyar, Lenske, Mosel \cite{SLM13} with 
Re~$a_{\eta N}=1.0$~fm. This wide range of values introduces considerable 
uncertainty into the evaluation of $\eta$-nuclear spectra, as shown very 
recently by Friedman, Gal, Mare\v{s} (FGM) \cite{FGM13} for $1s_{\eta}$ 
nuclear bound states. Calculated $1s_{\eta}$ binding energies in $^{208}$Pb, 
for example, range approximately between 10 and 30 MeV. Generally and as 
naively expected, the larger and hence more attractive Re~$a_{\eta N}$ is, the 
larger is the calculated binding energy of a given $1s_{\eta}$ nuclear state. 
In particular, the $\eta$-nuclear interaction generated from the Green-Wycech 
(GW) \cite{GW05} Re~$a_{\eta N} \sim 1.0$~fm amplitude is sufficiently 
strong to bind additional single-particle $\eta$ states in heavy nuclei, 
as shown in Sect.~\ref{sec:results} of the present work. Regarding 
Im~$a_{\eta N}$, most analyses result in a much narrower interval of values 
between 0.2 to 0.3 fm. Therefore, one might think that calculated widths of 
$\eta$-nuclear states should exhibit little model dependence. However, this 
expectation is not borne out in the very recent calculations by FGM that find 
widths of the $1s_{\eta}$ state in $^{208}$Pb ranging from a few MeV to about 
25 MeV, depending on the assumed $\eta N$ interaction model. 

\begin{figure}[htb] 
\begin{center} 
\includegraphics[width=0.48\textwidth]{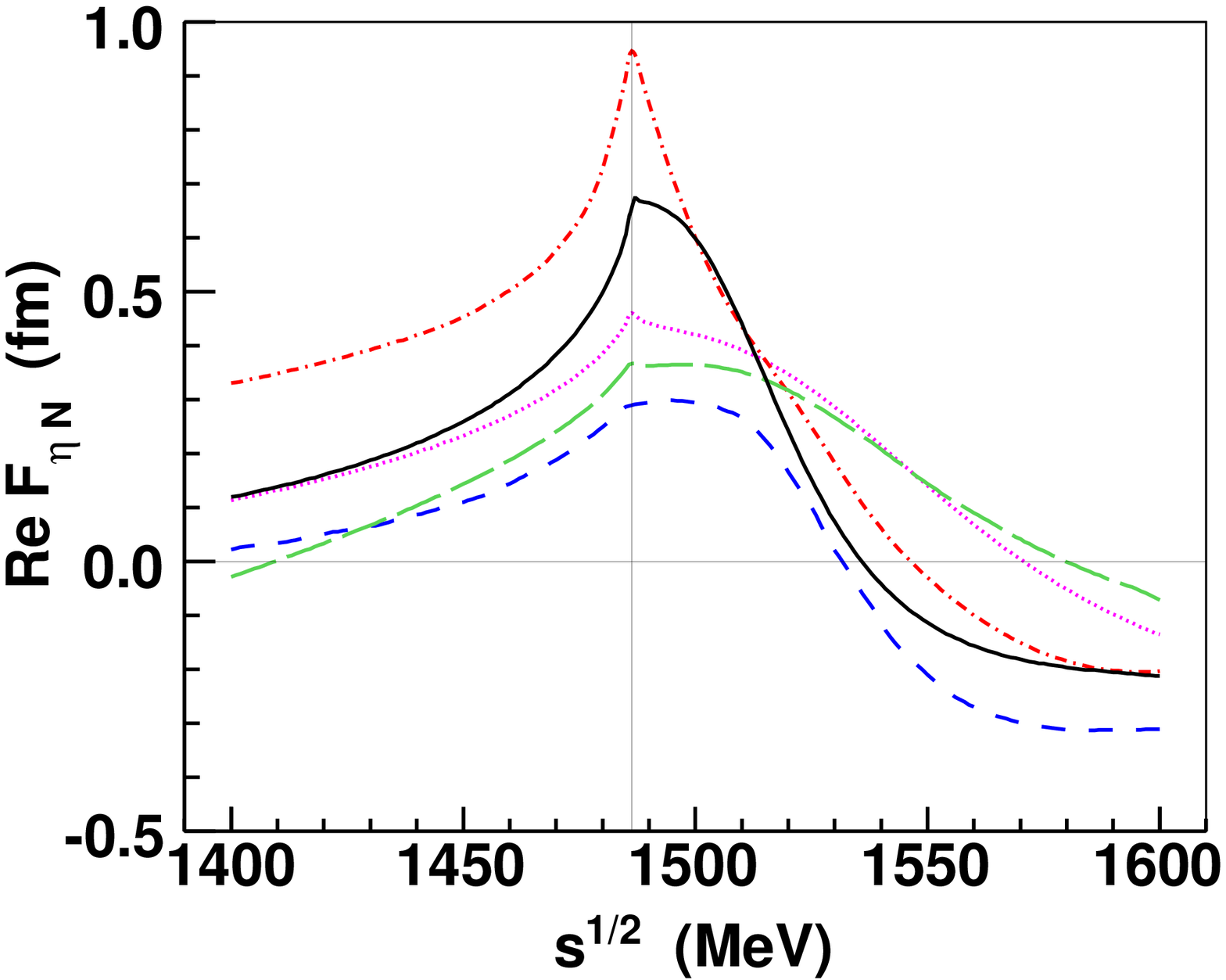} 
\includegraphics[width=0.48\textwidth]{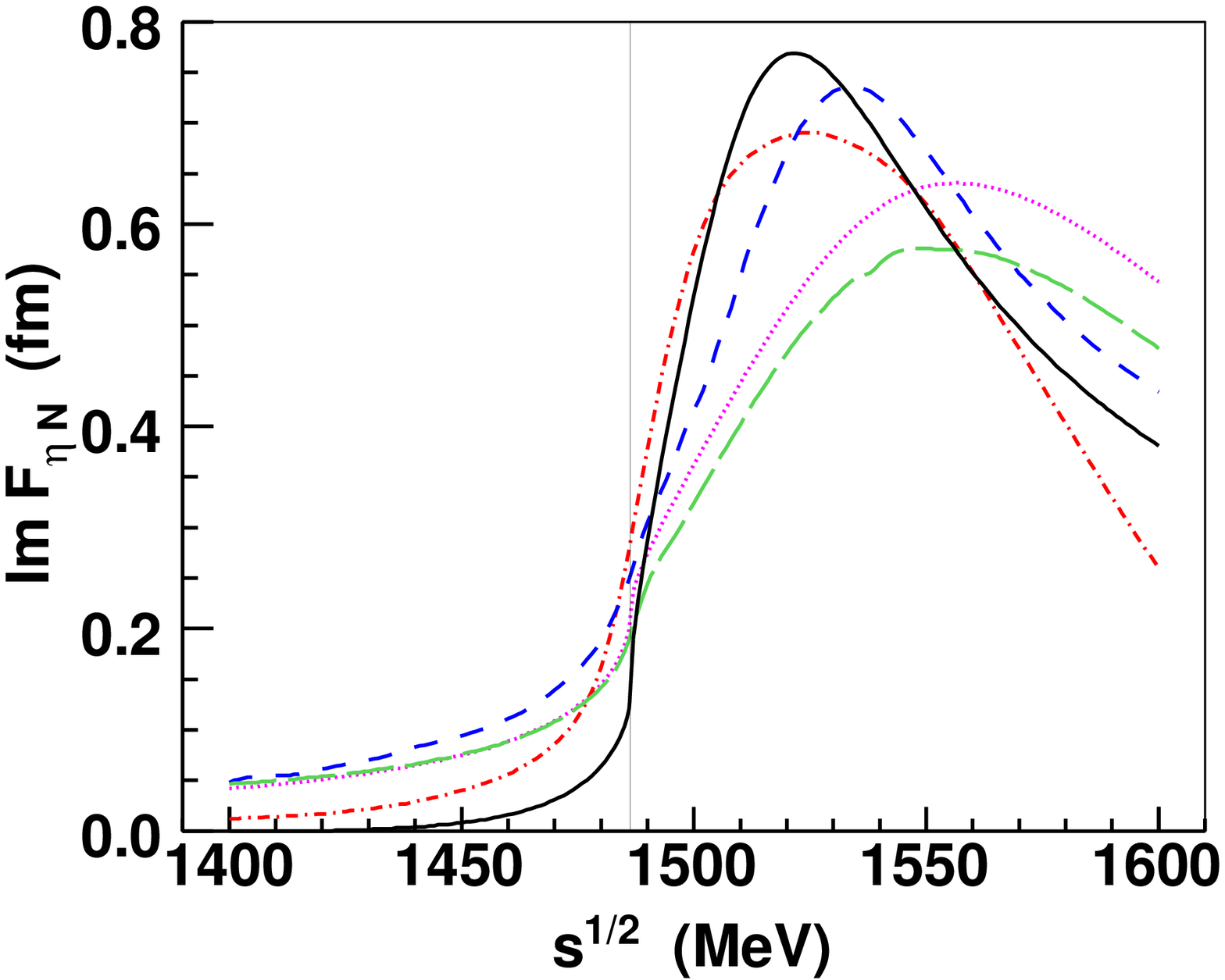} 
\caption{Real (left panel) and imaginary (right panel) parts of the $\eta N$ 
cm scattering amplitude $F_{\eta N}(\sqrt{s})$ as a function of the total cm 
energy $\sqrt{s}$ from five meson-baryon coupled-channel interaction models, 
in decreasing order of Re$\;a_{\eta N}$. Dot-dashed curves: GW \cite{GW05}; 
solid: CS \cite{CS13}; dotted: KSW \cite{KSW95}; long-dashed: M2 \cite{MBM12}; 
short-dashed: IOV \cite{IOV02}. The thin vertical line denotes the $\eta N$ 
threshold.} 
\label{fig:aEtaN1.eps} 
\end{center} 
\end{figure}

An important lesson of the FGM work is that the in-medium $\eta N$ scattering 
amplitudes that serve input in the calculation of $\eta$-nuclear bound states 
cannot be determined in terms of threshold $\eta N$ scattering amplitudes 
alone, be it free-space or in-medium threshold amplitudes. It was shown 
that $\eta N$ scattering amplitudes down to about 50 MeV below threshold 
are involved in $\eta$-nuclear bound state calculations \cite{FGM13}. 
In the coupled-channel studies of the $N^{\ast}(1535)$ resonance region 
cited above, the extrapolation from $\sqrt{s}\sim 1535$~MeV to the $\eta N$ 
threshold at $\sqrt{s_{\eta N}}=1487$~MeV and further down to the $\eta N$ 
subthreshold region introduces appreciable model dependence which is reflected 
in the large span of reported values for $a_{\eta N}$, particularly for its 
real part. The only model-independent property shared by such studies is that 
both real and imaginary parts of the $\eta N$ scattering amplitude decrease 
steadily as one goes below the $\eta N$ threshold. This is demonstrated in 
Fig.~\ref{fig:aEtaN1.eps} where the real and imaginary parts of the $\eta N$ 
center-of-mass (cm) scattering amplitude $F_{\eta N}(\sqrt{s})$ are plotted 
as a function of $\sqrt{s}$ for five different $s$-wave interaction models. 
The position of the $N^{\ast}(1535)$ resonance is closely related to the 
maximum of Im$\;F_{\eta N}(\sqrt{s})$ on the right panel. We note that no 
simple relationship emerges between the hierarchy of $\eta N$ scattering 
amplitudes shown here, neither for Re\;$F_{\eta N}$ nor for Im\;$F_{\eta N}$, 
and the gross properties of the $N^{\ast}(1535)$ resonance such as its peak 
position or width. 

In Ref.~\cite{FGM13}, in-medium $\eta N$ scattering amplitudes 
$F_{\eta N}(\sqrt{s},\rho)$ that satisfy the low-density requirement 
$F_{\eta N}(\sqrt{s},\rho)\to F_{\eta N}(\sqrt{s})$ upon density $\rho \to 0$ 
were formed from the corresponding free-space $\eta N$ scattering amplitudes 
$F_{\eta N}(\sqrt{s})$ using the Ericson-Ericson multiple-scattering 
reformulation given in Ref.~\cite{WRW97} (and employed recently also 
in Ref.~\cite{FG13}): 
\begin{equation} 
F_{\eta N}(\sqrt{s},\rho)=\frac{F_{\eta N}(\sqrt{s})}
{1+\xi(\rho)(\sqrt{s}/E_N)F_{\eta N}(\sqrt{s})\rho}\;,
\label{eq:WRW} 
\end{equation} 
where 
\begin{equation} 
\xi(\rho) = \frac{9\pi}{4p_F^2}I(\kappa), \;\; I(\kappa)=4\int_0^{\infty}
{\frac{dt}{t}\,\exp(-\kappa t)\,j_1^2(t)}\; . 
\label{eq:WRWFG} 
\end{equation} 
Here $E_N$ is the nucleon energy, often approximated by its mass $m_N$, 
$p_F$ is the local Fermi momentum corresponding to density 
$\rho=2p_F^3/(3\pi^2)$ and $\xi(\rho)$ accounts for Pauli blocking. 
At threshold, $\kappa=0$, $I(\kappa)=1$ and $\xi(\rho)=9\pi/(4p_F^2)$. 
For subthreshold energies represented by $\eta$ nuclear complex binding 
energies $B_{\eta}+{\rm i}\Gamma/2$, $\kappa=\sqrt{2m_{\eta}(B_{\eta}+
{\rm i}\Gamma/2)}/p_F$, $I(\kappa)$ remains dominantly real but with 
magnitude less than one, typically 0.5. 
We have used expression (\ref{eq:WRWFG}) to revise the FGM bound-state 
calculations done for $\kappa = 0$. This leads to a moderate increase 
of the $\eta$-nuclear attraction and, consequently, to binding energies 
that are somewhat larger, by a few MeV at most, than those reported by 
FGM \cite{FGM13}. The calculated widths hardly change. 

The primary aim of the present work is to extend the FGM analysis by 
using in-medium $\eta N$ interactions constructed here within the recent 
chirally-inspired meson-baryon coupled-channel model by Ciepl\'y and 
Smejkal (CS) \cite{CS13} in which self-energy insertions, disregarded 
by FGM \cite{FGM13}, are now included. This construction follows 
a similar one by these authors for in-medium $s$-wave $\bar K N$ 
interactions in Ref.~\cite{CS12} where several $\bar K$-nuclear 
applications are reviewed \cite{CFGK11,CFGGM11a,CFGGM11b,BGL12,GM12,FG12}. 
The corresponding in-medium $\eta N$ scattering amplitudes 
$F_{\eta N}(\sqrt{s},\rho)$ are applied in the present work 
directly within a comprehensive study of $\eta$-nuclear bound states, 
without having to approximate in-medium amplitudes by using the 
multiple-scattering expressions (\ref{eq:WRW}) and (\ref{eq:WRWFG}) 
(which nevertheless are found to provide a very good approximation to within 
few percent). Comparison is also made with another study done within 
a different coupled-channel approach \cite{IOV02} and its in-medium 
implementation \cite{IO02}, and with a different procedure of handling 
the energy dependence of in-medium $\eta N$ scattering amplitudes in 
$\eta$-nuclear bound-state calculations \cite{GR02,JNH02}. 

For a given $\eta N$ interaction model, the calculation of $\eta$-nuclear 
bound states in the present work follows the same steps introduced by FGM 
\cite{FGM13}. Thus, one solves the Klein-Gordon (KG) equation 
\begin{equation}
[\:\nabla^2 + {\tilde\omega}_{\eta}^2 - m_{\eta}^2 -
\Pi_{\eta}(\omega_{\eta},\rho)\:]\:\psi=0 \; , 
\label{eq:KG}
\end{equation}
where ${\tilde\omega}_{\eta}=\omega_{\eta}-{\rm i}\Gamma_{\eta}/2$ and
$\omega_{\eta}=m_{\eta}-B_{\eta}$, with $B_{\eta}$ and $\Gamma_{\eta}$
the binding energy and the width of the $\eta$-nuclear bound state,
respectively.
The self-energy operator $\Pi_{\eta}(\omega_{\eta},\rho)$ is given by 
\begin{equation}
\Pi_{\eta}(\omega_{\eta},\rho)\equiv 2\omega_{\eta}V_{\eta}=-4\pi
\frac{\sqrt{s}}{E_N}
F_{\eta N}(\sqrt{s},\rho)\rho \; ,
\label{eq:Pi}
\end{equation}
where $\rho$ is the nuclear density (normalized to the number of nucleons 
$A$) and 
\begin{equation} 
s=(\omega_{\eta}+E_N)^2-({\vec p}_{\eta}+{\vec p}_N)^2 
\label{eq:s} 
\end{equation} 
is the Lorentz invariant Mandelstam variable. The factor $(\sqrt{s}/E_N)$ 
transforms the in-medium cm scattering amplitude $F_{\eta N}(\sqrt{s},
\rho)$ to the corresponding laboratory (lab) amplitude which for $A\gg 1$ 
is the one relevant in $\eta$-nuclear calculations. 
In the lab system, in distinction from the $\eta N$ two-body cm system where 
${\vec p}_{\eta}+{\vec p}_N=0$, the two momenta are determined separately by 
the nuclear medium and their combined contribution is well approximated by 
a non-zero $(p_{\eta}^2+p_N^2)$ term. Including this negative-definite 
contribution to $s$ in the evaluation of $F_{\eta N}(\sqrt{s},\rho)$ weakens 
both real and imaginary parts of the $\Pi_{\eta}(\omega_{\eta},\rho)$ 
self-energy input to the KG equation, reducing thereby the calculated 
$\eta$-nuclear binding energy and width with respect to all other calculations 
that preceded FGM. The actual binding energy and width calculation requires 
a {\it self-consistent} procedure, since each of the four kinematical 
variables $\omega_\eta$, $E_N$, $p_{\eta}^2$ and $p_N^2$ of which $\sqrt{s}$ 
consists depends in the nuclear medium on the nuclear density $\rho$. In 
particular, as discussed below in Sect.~\ref{sec:SC}, $p_{\eta}^2$ depends 
on $\rho$ primarily through the self-energy $\Pi_{\eta}(\omega_{\eta},\rho)$, 
so even the input self-energy operator requires iterative cycles to become 
uniquely determined. Previous $\eta$-nuclear calculations were only concerned 
with the dependence of the input $\sqrt{s}$ (through the $\omega_{\eta}$ term) 
on the output binding energy $B_{\eta}$ \cite{GR02,JNH02}. Another improvement 
offered here, as well as in the recent FGM calculation, consists of solving 
RMF equations of motion for the in-medium nucleons in sequence and 
self-consistently \cite{MFG06} with the $\eta$-nuclear KG equation, thereby 
allowing the $\eta$ meson to polarize the nuclear core. However, the core 
polarization effect on $B_{\eta}$ and $\Gamma_{\eta}$ was found in these 
dynamical calculations to be smaller than 1~MeV, thus justifying the use of 
static nuclear densities at the present state of the art in $\eta$-nuclear 
calculations. 

The paper is organized as follows. In Sect.~\ref{sec:etaN} we derive the 
in-medium $\eta N$ scattering amplitudes from the free-space coupled-channel 
chirally-inspired separable-interaction NLO30$_{\eta}$ model due to 
CS \cite{CS13}. Our methodology of treating energy dependence, 
density dependence and combining these together self-consistently as 
in Ref.~\cite{FGM13} is outlined in Sect.~\ref{sec:SC}, while results 
of dynamical bound-state calculations of $\eta$-nuclear states are reported 
and discussed in Sect.~\ref{sec:results}. The paper ends with summary and 
outlook in Sect.~\ref{sec:sum}.

\section{In-medium chirally-inspired coupled-channel $\eta N$ interaction} 
\label{sec:etaN}

It was made clear in the Introduction that in-medium $\eta N$ scattering 
amplitudes $F_{\eta N}(\sqrt{s},\rho)$ are needed for constructing the $\eta$ 
self-energy operator $\Pi_{\eta}(\omega_{\eta},\rho)$, or equivalently the 
$\eta$-nuclear potential (\ref{eq:Pi}). In close relationship to our 
recent works on $\bar{K}$-nuclear interaction \cite{CFGGM11a,CFGGM11b} 
we employ chirally motivated meson-baryon $s$-wave scattering amplitudes 
$F_{ij}$, given in the two-body cm system by a separable form 
\begin{equation} 
F_{ij}(k,k';\sqrt{s})=g_{i}(k^{2}) \: f_{ij}(\sqrt{s}) \: g_{j}(k'^{2}) \; , 
\label{eq:Fsep} 
\end{equation} 
with off-shell form factors chosen as 
\begin{equation} 
g_{j}(k)=1/[1+(k/ \alpha_{j})^2] \; ,  
\label{eq:FF} 
\end{equation} 
superposed on the purely energy-dependent {\it reduced} amplitudes $f_{ij}$. 
The indices $i$ and $j$ label meson-baryon coupled channels: $\pi N$, 
$\eta N$, $K\Lambda$ and $K\Sigma$, in order of their threshold energies. 
The meson-baryon cm momenta in the initial (final) state are denoted $k$ 
($k'$), $\sqrt{s}$ stands for the total energy in the two-body cm system, 
and the inverse-range parameters $\alpha$ characterize the interaction range 
in the specified meson-baryon channels. The scattering amplitudes $F_{ij}$ 
solve the coupled-channels Lippmann-Schwinger (LS) equation 
\begin{equation} 
F = V + V G F \; ,
\label{eq:LSE} 
\end{equation} 
where $G$ stands for the intermediate-state Green's function and the 
coupled-channels potential matrix $V$ is given in separable form 
\begin{equation} 
V_{ij}(k,k';\sqrt{s})=g_{i}(k^{2}) \: v_{ij}(\sqrt{s}) \: g_{j}(k'^{2})  \; , 
\label{eq:Vsep} 
\end{equation} 
with the same form factors $g_{j}(k^{2})$ as in (\ref{eq:Fsep}), here given by 
Eq.~(\ref{eq:FF}). 
The energy-dependent $v_{ij}(\sqrt{s})$ elements of the potential matrix are 
determined by matching to SU(3) chiral amplitudes derived to a given order 
of the chiral expansion. While the basic features of the $\bar{K}N$ coupled 
channel interactions are satisfactorily described already by the leading 
order (LO) Tomozawa-Weinberg (TW) term, a good reproduction of the $\pi N$ and 
$\eta N$ experimental data requires next-to-leading-order (NLO) contributions. 
In the present work we use model NLO30$_{\eta}$ from the recent work of 
Ciepl\'{y} and Smejkal \cite{CS13}. 

The intermediate-state meson-baryon Green's function $G$ is diagonal in the 
channel indices $i,j$. It follows then that the LS equations (\ref{eq:LSE}) 
allow for algebraic solution, with reduced amplitudes $f_{ij}(\sqrt{s})$ given 
by  
\begin{equation} 
f_{ij}(\sqrt{s}, \rho)=\left[ (1 - v(\sqrt{s}) \cdot G(\sqrt{s}, \rho))^{-1} 
\cdot v(\sqrt{s}) \right]_{ij} \; .
\label{eq:fij} 
\end{equation} 
Marked in this equation explicitly, in addition to energy dependence, is 
also a density dependence of the reduced amplitudes implied by the density 
dependence that the Green's function acquires in the nuclear medium owing 
to (i) Pauli blocking and to (ii) self-energy insertions. The Green's 
function in channel $n$ is expressed as 
\begin{equation} 
G_{n}(\sqrt{s},\rho) = -4\pi \: \int_{\Omega_{n}(\rho)}
\frac{d^{3}p}{(2\pi)^{3}}\frac{g_{n}^{2}(p^{2})}
{k_{n}^{2}-p^{2} -\Pi^{(n)}(\sqrt{s},\rho) +{\rm i}0} \; , 
\label{eq:Grho} 
\end{equation} 
where the integration over the intermediate meson-baryon momenta is 
restricted to a region $\Omega_{n}(\rho)$ ensuring that in channels 
involving nucleons the intermediate nucleon energy is above the Fermi 
level (see Ref.~\cite{WKW96} for details). The self-energy 
$\Pi^{(n)}(\sqrt{s},\rho)$ stands for the {\it sum} of hadron 
self-energies $\Pi^{(n)}_{h}(\sqrt{s},\rho)$ in channel $n$, 
given in terms of hadron-nucleus potentials $V_{h}(\sqrt{s},\rho)$: 
\begin{equation} 
\Pi^{(n)}_{h}(\sqrt{s},\rho) = 2\mu_{n}(\sqrt{s})V_{h}(\sqrt{s},\rho) \;.   
\label{eq:SE} 
\end{equation} 
Here $\mu_{n}(\sqrt{s})$ is the meson-baryon relativistic reduced energy 
and the potential $V_{h}$ is chosen for simplicity linear in the density:  
\begin{equation} 
V_{h}(\sqrt{s},\rho)=v_{h}(\sqrt{s})\rho / \rho_{0} \; , 
\label{eq:V} 
\end{equation} 
except for $V_{\eta}$ which is determined self-consistently as explained 
below. The value $\rho_{0}=0.17$~fm$^{-3}$ is used for nuclear-matter density. 
%{\footnote{the self-energies that enter Eq.~(\ref{eq:Grho}) differ by 
%a factor $\mu_n/E_h$, where $E_h$ denotes the hadron energy in the 
%meson-baryon cm, from the self-energies commonly used in 
%the equation of motion for the dressed hadron in nuclear matter. 
%This factor originates from a nonrelativistic treatment of our 
%intermediate-state Green's function.}}
In Table~\ref{tab:depth} we list the self-energy (SE) input hadron potential 
depths $v_{h}$ at the $\eta N$ threshold. The baryon nuclear-matter potentials 
are the same ones used in our earlier work \cite{CFGM01}, with energy 
dependence disregarded. The meson potential depths are determined as 
itemized below. 

\begin{table}[hbt]
\begin{center}
\caption{Potential depths $v_{h}$ (in MeV), Eq.~(\ref{eq:V}) at 
the $\eta N$ threshold $\sqrt{s_{\eta N}}=1487$ MeV, providing input to 
self-energies in Eqs.~(\ref{eq:Grho}) and (\ref{eq:SE}), with values from 
Ref.~\cite{CFGM01} for baryons and values discussed in the text for mesons.} 
\begin{tabular}{ccccc}
\noalign{\smallskip}\hline\noalign{\smallskip}
$\pi$ & $K$ & $N$ & $\Lambda$ & $\Sigma$ \\
\noalign{\smallskip}\hline\noalign{\smallskip}
20$-$i40 & 31.6 & $-$(60+i10) & $-$(30+i10) & 30$-$i10 \\ 
\hline
\end{tabular}
\label{tab:depth}
\end{center}
\end{table} 

\begin{itemize} 
\item For pions we derived empirical $\pi N$ scattering amplitudes from 
SAID \cite{gwdac} including several partial waves beyond $s$ waves. With 
on-shell pion momenta of over 400 MeV/c at the $\eta N$ threshold region, 
Pauli blocking is negligible and the $\pi N$ free-space amplitude 
$F_{\pi N}(\sqrt{s})$ should approximate well the $\pi N$ in-medium 
amplitude. The corresponding $\pi N$ free-space forward scattering 
amplitude was then substituted in 
\begin{equation} 
2\mu_{\pi N}(\sqrt{s})v_{\pi}(\sqrt{s})=-4\pi F_{\pi N}(\sqrt{s};0^{\circ})
\rho_{0}   
\label{eq:depth} 
\end{equation}  
to estimate the pion-nuclear potential depth. The resulting pion potential 
and SE (with $v_{\pi}$ listed in Table~\ref{tab:depth}) are weakly 
repulsive but substantially absorptive, with little energy dependence around 
the $\eta N$ threshold, exercising negligible effect on the outcome in-medium 
$\eta N$ scattering amplitude in the subthreshold region of interest. Thus, 
reversing the sign of the real part of the pion SE or setting it 
to zero affects marginally the resulting $\eta N$ amplitudes. Reducing the 
imaginary part also has marginal effect. It appears that the pion SE is 
insignificant owing to the considerably larger kinetic energy of pions at 
the $\eta N$ threshold region. Furthermore, we recall that the $\pi N$ and 
$\eta N$ systems are decoupled at LO, both communicating with each other 
through the attractive kaon-hyperon systems which generate the two major 
$(1/2)^-$ $N^{\ast}(1535)$ and $N^{\ast}(1650)$ resonances above the $\eta N$ 
threshold. 
\item For kaons we used a value of $v_K=30$~MeV at the $KN$ threshold 
$\sqrt{s_{KN}}=1433$~MeV, averaging over two recent phenomenological 
derivations from GSI experiments: (i) $(20\pm 5)$~MeV from the in-medium 
$K^0$ inclusive cross sections in $\pi^-$-induced reactions on several 
nuclear targets at 1.15 GeV/c \cite{FOPI09}, and (ii) about 40 MeV from 
transverse momentum spectra and rapidity distributions of $K^0_s$ in 
Ar+KCl reactions at a beam kinetic energy of 1.756~$A$~GeV \cite{HADES10}. 
This value of $v_K$ is very close to the value $v_K\approx 32.1$~MeV derived 
from the energy-independent SE employed by Inoue and Oset \cite{IO02}. To 
account for energy dependence we multiplied the $KN$-threshold value of $v_K$ 
by the ratio of reduced energies $\mu_{KN}(\sqrt{s})/\mu_{KN}(\sqrt{s_{KN}})$ 
which provides a good approximation away from the $KN$ threshold, resulting in 
the tabulated value. 
%$\Pi^{K}(\rho)=0.13\: m_{K}^{2}\: \rho / \rho_{0}$ \cite{IO02}. 
The kaon potential and SE are moderately repulsive in the $\eta N$ 
threshold, exercising a nonnegligible effect on the resulting in-medium 
$\eta N$ scattering amplitude in the subthreshold region of interest, 
as shown below. 
\item The self-consistently derived $\Pi_{\eta}$ is largely independent of 
whatever input $v_{\eta}$ value is used. For a representative value, consider 
implementing multi-channel Pauli blocking without introducing simultaneously 
any SE. This gives $v_{\eta}$=$-(42.1+{\rm i}16.1)$~MeV at the $\eta N$ 
threshold in model NLO30$_{\eta}$ of CS \cite{CS13}. The LS equations 
were then solved iteratively to achieve convergence for the in-medium SE 
$\Pi_{\eta}$ of Eqs.~(\ref{eq:KG}) and (\ref{eq:Pi}). For completeness 
we note that $\Pi_{\eta}$ is related to the $\eta N$-channel SE 
$\Pi^{(\eta N)}_{\eta}$ by $\Pi_{\eta}=(\sqrt{s}/E_N)\Pi_{\eta}^{(\eta N)}$. 
We recall that $\Pi_{\eta}$ is the only SE constrained by a self-consistency 
requirement in our iterative solution of the LS equations. 
\end{itemize} 

With self-energies accounted for, and considering the energy and density 
dependence of $\Pi_{\eta}$, the coupled-channels LS equations are solved 
iteratively to achieve self-consistency, see Ref.~\cite{CFGGM11b} for details. 
No more than 5--7 iterations are normally needed to achieve the required 
precision. 

\begin{figure}[htb] 
\begin{center} 
\includegraphics[width=0.48\textwidth]{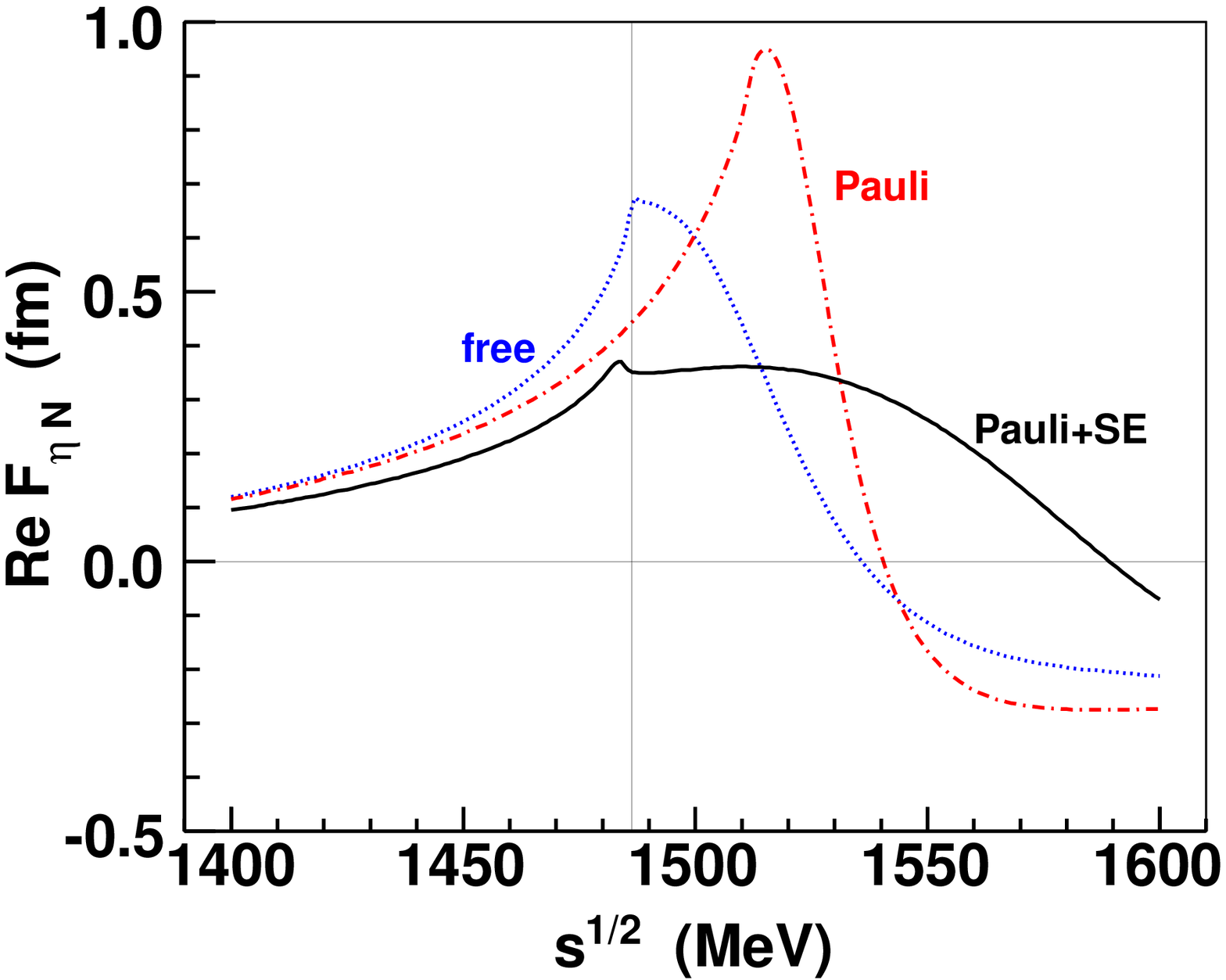} 
\includegraphics[width=0.48\textwidth]{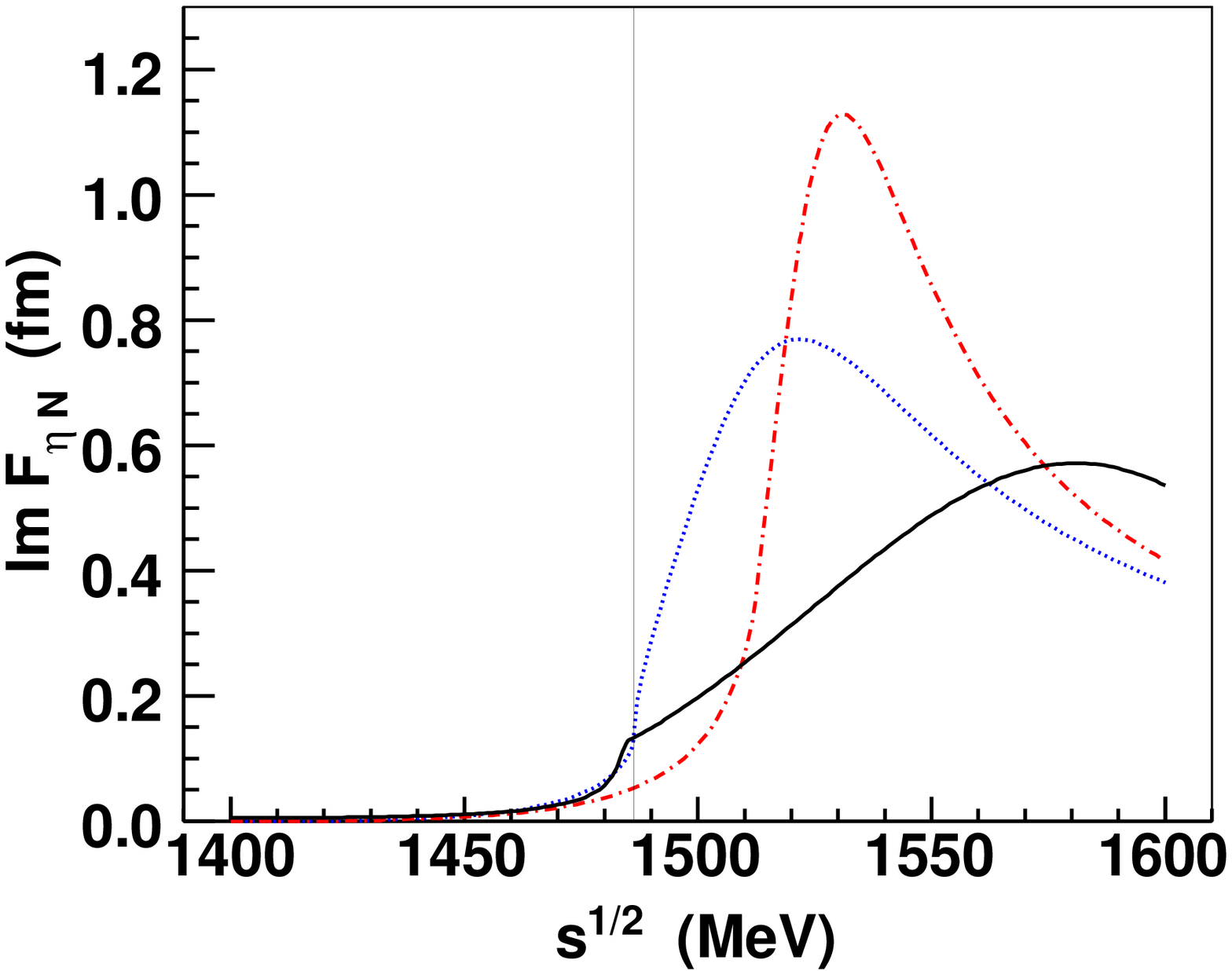} 
\caption{Real (left panel) and imaginary (right panel) parts of the 
$\eta N$ cm scattering amplitude generated in the NLO30$_{\eta}$ model 
of CS \cite{CS13}. Dotted curves: free-space amplitude (from Fig.~4 
in \cite{CS13}, same as solid curves in the preceding figure); dot-dashed: 
Pauli blocked in-medium amplitude for $\rho_0=0.17$~fm$^{-3}$; solid: 
including hadron self-energies in the Pauli blocked in-medium amplitude.} 
\label{fig:aEtaNrho} 
\end{center} 
\end{figure} 

The nuclear medium effect on the energy dependence of the $\eta N$ scattering 
amplitude is demonstrated in Fig.~\ref{fig:aEtaNrho}. With inverse-range 
parameter $\alpha_{\eta N}=1635$~MeV/c \cite{CS13}, any explicit momentum 
dependence is negligible in this model, and there is practically no 
difference between the amplitudes $F_{\eta N}$ and their respective reduced 
parts $f_{\eta N}$ (\ref{eq:Fsep}); hence, the self-energy $\Pi_{\eta}$ 
(\ref{eq:Pi}) has no explicit momentum dependence beyond the implicit one 
arising from the dependence of $s$, Eq.~(\ref{eq:s}), on $p_{\eta}$. We note 
that the peak structure observed in the figure for Im\;$F_{\eta N}$ may be 
ascribed to the $N^{\ast}(1535)$ resonance generated dynamically in this 
coupled-channel model. In-medium Pauli blocking (dot-dashed curves) shifts 
the resonance to higher energies, making it more pronounced. Implementing 
hadron self-energies (solid curves) spreads the resonance structure over 
a broad interval of energies, practically dissolving it in the nuclear medium. 
This behavior is different from that observed for the $\bar{K}N$ system where 
the hadron self-energies compensate to large extent for the effect of Pauli 
blocking and bring the peak structure back below the $\bar{K}$ threshold, 
resulting in strong in-medium attraction with little energy dependence at 
subthreshold energies relevant for kaonic atoms and for $K^-$-nuclear bound 
states \cite{CFGGM11b}. For the $\eta N$ system, in contrast, the in-medium 
amplitudes decrease rapidly in going to the subthreshold energies relevant 
for $\eta$-nuclear bound states and are weaker than the respective free-space 
amplitudes. In particular, the relatively large value of the free-space 
Re\;$a_{\eta N}$ is almost halved for nuclear matter density. 
%Finally, the solid curves in Fig.~\ref{fig:aEtaNrho} show the effect 
%of reversing the real part of the $\Sigma$-hyperon self-energy in the 
%calculation of in-medium $F_{\eta N}$. Although the $KY$ channels are closed 
%in the $\eta N$ threshold region, they do exercise substantial impact on the 
%in-medium $F_{\eta N}$ through the nearby $N^{\ast}(1535)$ resonance which 
%is generated already at LO by these two-body channels. This reversal pulls 
%back the $N^{\ast}(1535)$ resonance, raising the in-medium $F_{\eta N}$ in 
%the subthreshold region almost to the level of its free-space counterpart 
%for the real part, and considerably above it for the imaginary part. 
%This behavior of the in-medium $F_{\eta N}$ bears some similarity to that 
%of the in-medium $F_{\eta N}$ constructed by Inoue and Oset \cite{IO02} who 
%used outdated estimates for $\;V_0^{\Sigma}$ (for a comprehensive discussion 
%see Ref.~\cite{FG07})

\begin{figure}[htb] 
\begin{center} 
\includegraphics[width=0.48\textwidth]{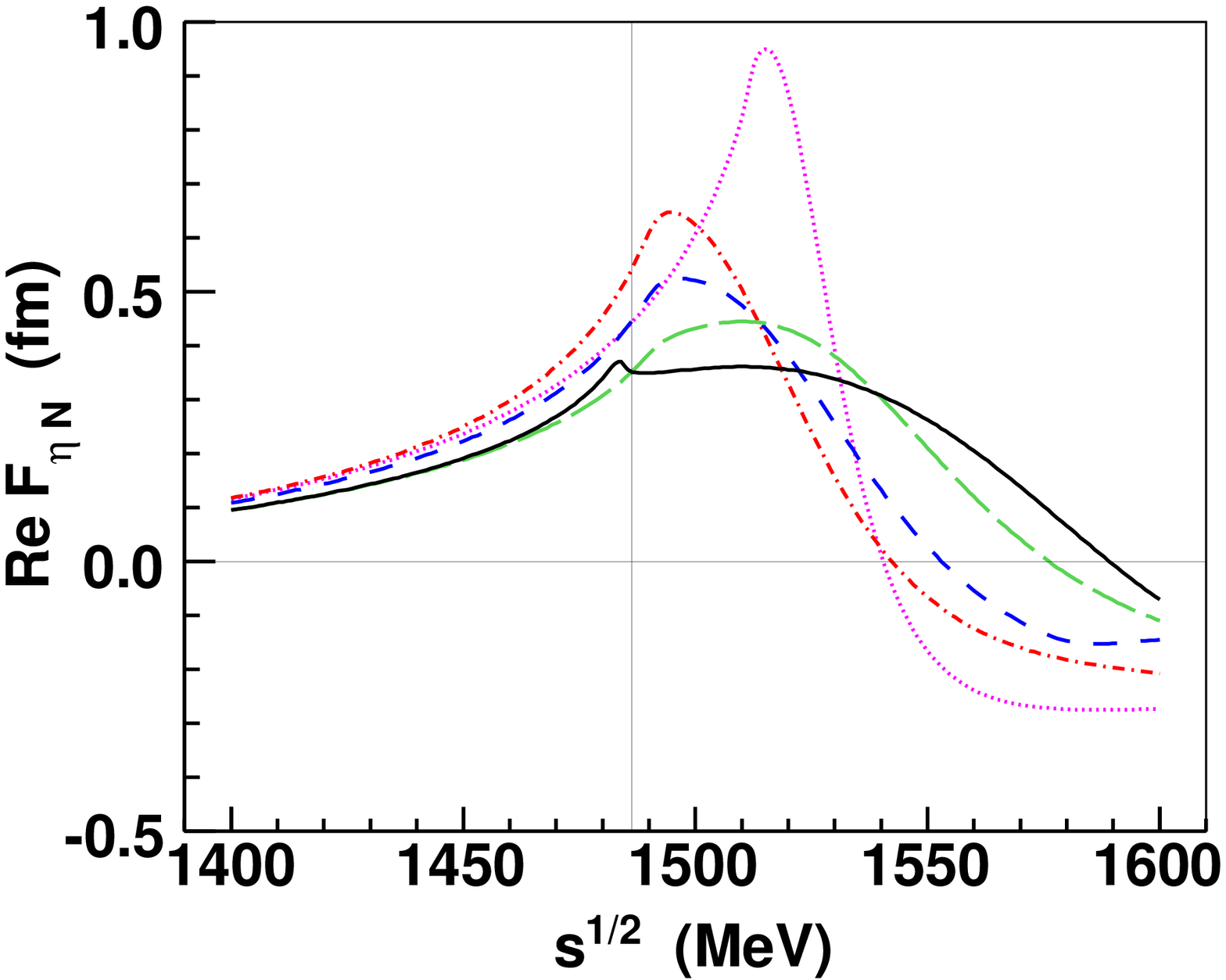} 
\includegraphics[width=0.48\textwidth]{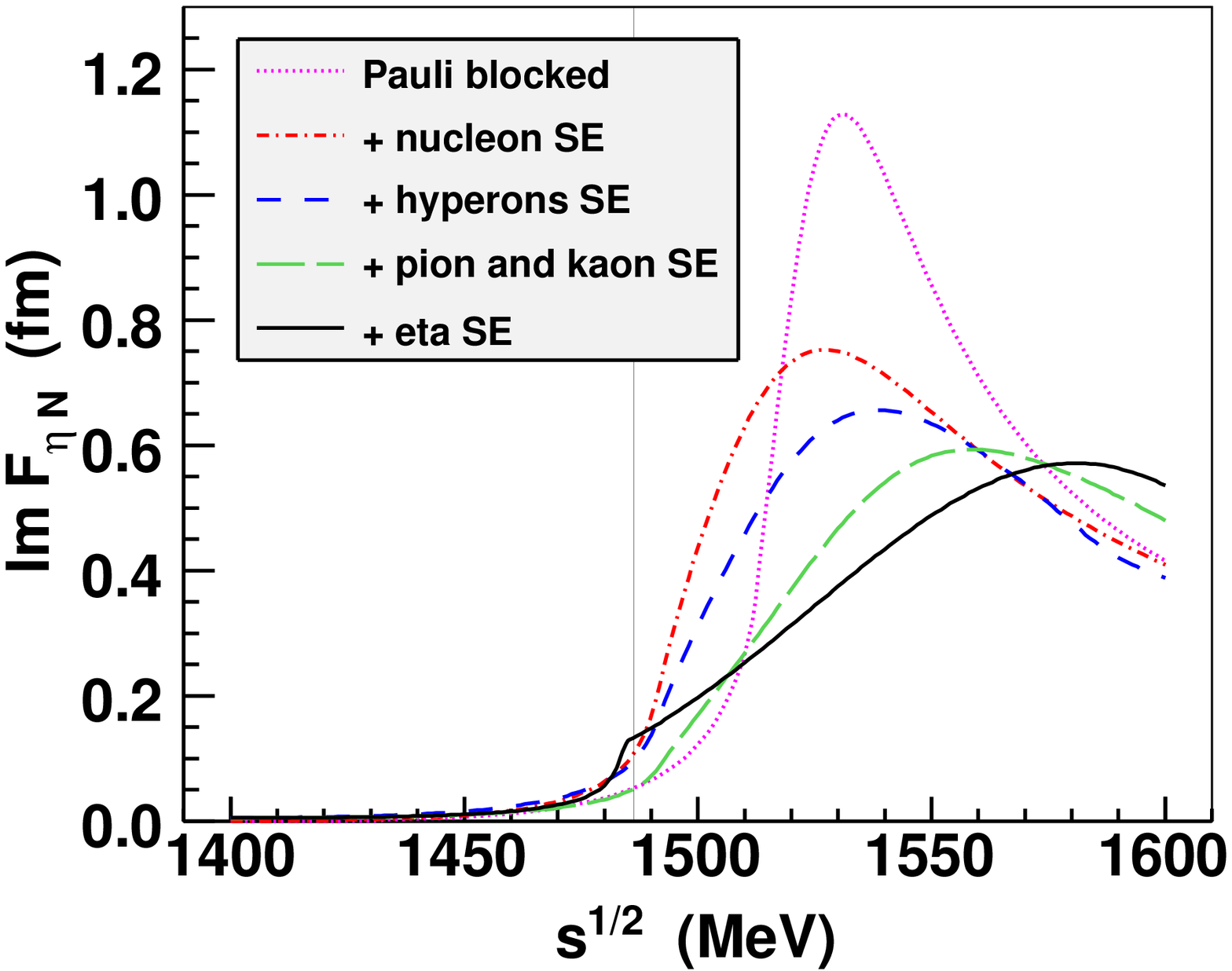} 
\caption{Effects of introducing self-energies on the real (left panel) and 
imaginary (right panel) parts of the Pauli-blocked $\eta N$ cm scattering 
amplitude generated in the NLO30$_{\eta}$ model of CS \cite{CS13} for 
$\rho_0=0.17$~fm$^{-3}$. Dot-dashed curves: before adding self-energies. 
Self-energies are added sequentially to nucleons (dotted), to hyperons 
(short-dashed), to pions and kaons (long-dashed), and self-consistently 
to the $\eta$ meson (solid).} 
\label{fig:SEtest} 
\end{center} 
\end{figure} 

The sensitivity of the in-medium $\eta N$ cm scattering amplitude $F_{\eta N}$ 
to various SE insertions is demonstrated in Fig.~\ref{fig:SEtest}. Dot-dashed 
curves show the in-medium Pauli-blocked amplitude, without any self-energy 
insertion. We then introduce successively self-energies due to nucleons, 
hyperons, mesons (excluding $\eta$), and finally in solid curves also the 
$\eta$ SE self-consistently. In all the cases displayed here the amplitudes 
decrease monotonically in going deeper below the $\eta N$ threshold. The 
effect of adding self-energies to the Pauli blocked amplitude is seen to be 
moderate at best. The resulting Re\;$F_{\eta N}$ is somewhat weaker than its 
Pauli-blocked counterpart. This is not the case for Im\;$F_{\eta N}$ around 
the $\eta N$ threshold, but its two versions (with and without self-energies) 
become equally weak about 20 MeV below threshold. 

%Whereas the energy dependence of the assumed kaon self-energy matters little 
%at the $\eta N$ threshold and below, the omission of $V_0^K$ shown by dotted 
%curves results in substantially incresed values of both real and imaginary 
%parts of the in-medium $F_{\eta N}$. 

\section{In-medium energy and density dependence}
\label{sec:SC}

The methodology of calculating {\it self-consistently} binding energies 
and widths of $\eta$-nuclear states has been presented recently by FGM 
\cite{FGM13}. Two novelties of this study are the derivation of the 
$\eta$-nuclear potential from in-medium $\eta$-nucleon scattering amplitudes 
at subthreshold energies, and the introduction of Relativistic Mean Field 
(RMF) equations for nucleons that are solved dynamically along with the KG 
equation (\ref{eq:KG}) for the $\eta$ meson, allowing thus for polarization 
of the nucleus by the bound meson. This approach was applied beforehand 
in analyses of kaonic atoms data \cite{FG13} and in calculations of strongly 
bound $K^-$-nuclear states \cite{GM12}. Here we outline the methodology 
of handling self-consistently in-medium $\eta N$ subthreshold scattering 
amplitudes $F_{\eta N}(\sqrt{s},\rho)$ for use in $\eta$-nuclear bound-state 
calculations. 

We recall that the meson-baryon Mandelstam variable $s$ is given by 
$s=(m_{\eta}+m_N-B_{\eta}-B_N)^2-({\vec p}_{\eta}+{\vec p}_N)^2$, 
including a non-zero in-medium momentum dependent term that provides 
additional downward energy shift to that arising from the sum of binding 
energies $(B_{\eta}+B_N)$. Near threshold, to leading order in binding and 
kinetic energies with respect to masses, one may approximate $\sqrt{s}$ 
by \cite{FGM13,CFGGM11a,CFGGM11b} 
\begin{equation}
\sqrt{s} \approx m_{\eta}+m_N - B_N - B_{\eta} - \xi_N\frac{p_N^2}{2m_N}
- \xi_{\eta}\frac{p_{\eta}^2}{2m_{\eta}} \; ,
\label{eq:approx}
\end{equation}
where $\xi_{N(\eta)}\equiv m_{N(\eta)}/(m_N+m_{\eta})$. To transform momentum 
dependence into density dependence, the nucleon kinetic energy $p_N^2/(2m_N)$ 
is approximated within the Fermi gas model by $T_N(\rho/\rho_0)^{2/3}$, with 
average bound-nucleon kinetic energy $T_N=23.0$ MeV at nuclear-matter density 
$\rho_0$. Furthermore, the ${\eta}$ kinetic energy $p_{\eta}^2/(2m_{\eta})$ is 
substituted within the local density approximation by $-B_{\eta}-{\rm Re}\:
V_{\eta}(\sqrt{s},\rho)$. Hence, the {\it in-medium} $\sqrt{s}=m_{\eta}+m_N+
\delta\sqrt{s}$ energy argument of $F_{\eta N}(\sqrt{s},\rho)$ in expression 
(\ref{eq:Pi}) for the self-energy exhibits explicit density 
dependence, with a form adjusted to respect the low-density limit, 
$\delta\sqrt{s}\to 0$ upon $\rho\to 0$, as used recently in $K^-$-atom 
studies \cite{FG13}:
\begin{equation}
\delta\sqrt{s}\approx -B_N\frac{\rho}{{\bar\rho}}-
\xi_N B_{\eta}\frac{\rho}{\rho_0}-
\xi_N T_N(\frac{\rho}{\rho_0})^{2/3}+
\xi_{\eta}{\rm Re}~V_{\eta}(\sqrt{s},\rho) \; .
\label{eq:sqrts}
\end{equation}
Here $B_N\approx 8.5$~MeV is an average nucleon binding energy and $\bar\rho$ 
is the average nuclear density. We note that in contrast with the assumption 
$p_{\eta}=0$ made normally in nuclear matter calculations, $p_{\eta}\neq 0$ 
in finite nuclei which explains the origin of $V_{\eta}$ in expression 
(\ref{eq:sqrts}). Furthermore, for an attractive $V_{\eta}$ and as long as 
$\rho\neq 0$, the shift of the two-body energy away from threshold is negative 
definite, $\delta\sqrt{s}<0$, even as $B_{\eta}\to 0$. 

It is clear from Eq.~(\ref{eq:sqrts}) that $\sqrt{s}$ depends on 
${\rm Re}~V_{\eta}(\sqrt{s},\rho)$ which by Eq.~(\ref{eq:Pi}) depends on 
$\sqrt{s}$. Therefore, for a given value of $B_{\eta}$, $V_{\eta}$ is 
determined {\it self-consistently} by iterating Eq.~(\ref{eq:sqrts}) with 
input from Eq.~(\ref{eq:Pi}). Up to six iterations suffice for convergence. 
This is done at each radial point where $\rho$ is given and for each 
$B_{\eta}$ value during the calculation of bound states. 

\begin{figure}[htb]
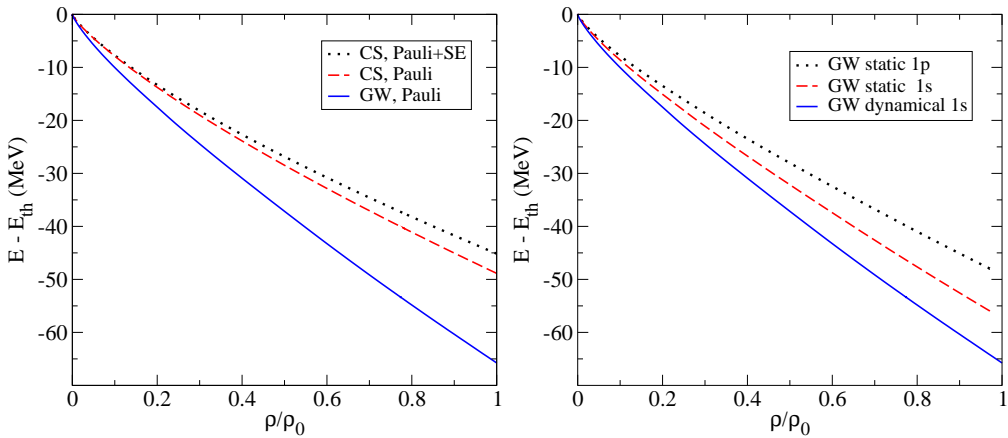
 
\begin{center} 
\includegraphics[width=0.48\textwidth]{srhocs2.eps} 
\includegraphics[width=0.48\textwidth]{rhotodelta1s1p.eps} 
\caption{Downward energy shift as a function of the nuclear density, 
obtained self-consistently using models GW \cite{GW05} and CS \cite{CS13} 
for in-medium $\eta N$ scattering amplitudes in $1s_{\eta}$ bound state 
dynamical calculations for Ca (left panel) and comparison with static 
calculations of the $1s_{\eta}$ and $1p_{\eta}$ bound states in Ca within 
model GW (right panel).} 
\label{fig:srhocs} 
\end{center} 
\end{figure} 

The downward subthreshold energy shift $\delta\sqrt{s}\equiv E-E_{\rm th}$ is 
plotted in Fig.~\ref{fig:srhocs} as a function of the nuclear density $\rho$ 
in Ca, evaluated self-consistently according to Eq.~(\ref{eq:sqrts}) for 
several different calculations. In the left panel of the figure we compare 
the correlation between $\delta\sqrt{s}$ and $\rho$ obtained by using the 
in-medium GW amplitudes $F_{\eta N}(\sqrt{s},\rho)$ with that for two 
versions of the in-medium CS amplitudes, all for the $1s_{\eta}$ state 
in Ca in dynamical calculations. The GW curve was calculated using 
Eqs.~(\ref{eq:WRW}) and (\ref{eq:WRWFG}), modifying thereby the FGM 
calculation that used $\kappa=0$ in Eq.~(\ref{eq:WRWFG}) and thus yielding 
up to 10 MeV larger energy shifts than shown in Fig.~2 of FGM \cite{FGM13}. 
The two CS versions account for Pauli blocking within a coupled-channel 
calculation, one version (Pauli) excludes and the other one (Pauli+SE) 
includes self-energies. It is seen that downward energy shifts ranging within 
55$\pm$10 MeV are correlated with nuclear central densities, and that the 
shift for the GW model exceeds that for the CS model, reflecting the stronger 
real part of the free-space amplitude $F_{\eta N}(\sqrt{s})$ in the GW model. 
Among the two CS versions, somewhat larger values of the downward energy 
shift are obtained in the version without self-energies. This reflects the 
larger values of Re~$F_{\eta N}(\sqrt{s},\rho)$ generated, on average, 
at subthreshold energies and finite densities upon suppressing the hadron 
self-energies. 

In the right panel of Fig.~\ref{fig:srhocs} we show similar results obtained 
for the GW in-medium amplitude within (i) a dynamical calculation of the 
$1s_{\eta}$ state in Ca (lowest curve, identical with the GW curve in the 
left panel) and within (ii) a static-density calculations of the $1s_{\eta}$ 
and $1p_{\eta}$ states in Ca. One observes that somewhat larger downward 
energy shifts are reached in the dynamical calculation, and that among the 
two static-density calculations larger downward energy shifts are obtained 
for the deeper bound state, $1s_{\eta}$. The difference of almost 10 MeV 
between the energy shifts at $\rho_0$ for these two GW static-density 
calculations is correlated with the slightly larger binding energy 
difference between the $1s_{\eta}$ and $1p_{\eta}$ states in Ca as 
demonstrated for dynamical calculations in Fig.~8 of the next section.

\section{Results and discussion} 
\label{sec:results} 

\begin{figure}[htb] 
\begin{center} 
\includegraphics[width=0.5\textwidth]{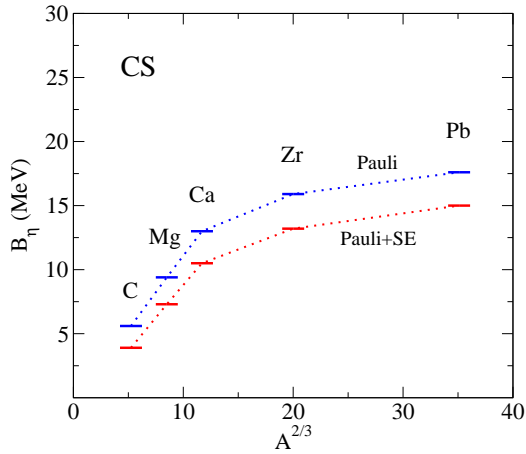} 
\caption{$1s_{\eta}$ binding energies in nuclei, calculated self-consistently 
and dynamically using in-medium $\eta N$ subthreshold scattering amplitudes 
constructed with (Pauli+SE) and without (Pauli) self energies in model 
NLO30$_{\eta}$ of CS \cite{CS13}. Pauli blocking is included within full 
coupled-channel calculations.} 
\label{fig:b-ales} 
\end{center} 
\end{figure} 

We have used the in-medium scattering amplitudes $F_{\eta N}(\sqrt{s},\rho)$ 
evaluated in model NLO30$_{\eta}$, as outlined in Sect.~\ref{sec:etaN}, 
within dynamical calculations of $\eta$-nuclear binding energies and 
widths in several nuclei across the periodic table, as described in 
Sect.~\ref{sec:SC}. Calculated binding energies in this model, marked CS, 
are shown in Fig.~\ref{fig:b-ales} for $1s_{\eta}$ nuclear states. The effect 
of including hadron self-energies (Pauli+SE) is demonstrated, resulting in 
2--3 MeV lower binding energies than those calculated with Pauli blocking 
only. The present procedure of treating Pauli blocking within in-medium 
coupled channels gives binding energies larger by 0.5--0.7 MeV than 
those calculated using the multiple-scattering approach specified by 
Eqs.~(\ref{eq:WRW}) and (\ref{eq:WRWFG}). Not shown in the figure are 
the remarkably small widths of about 2 MeV calculated for $1s_{\eta}$ 
nuclear states in model CS (these widths are shown below in Fig.~7). 

\begin{figure}[htb]
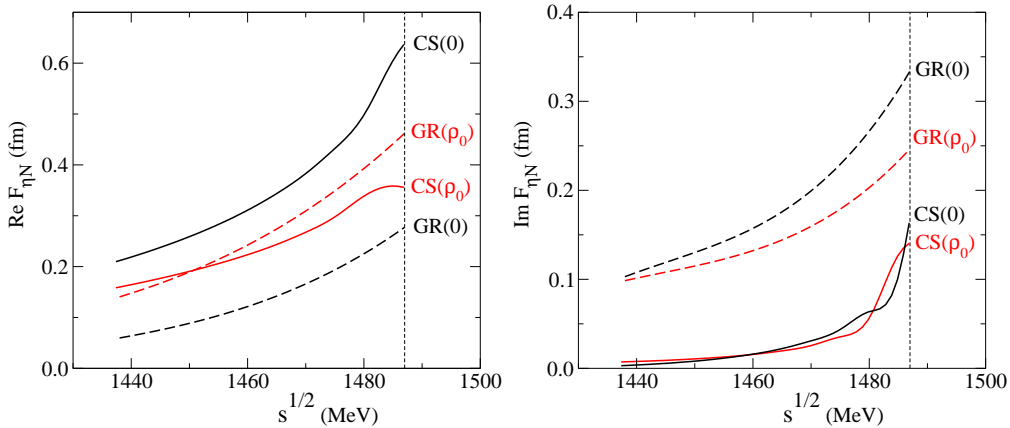
 
\begin{center} \includegraphics[width=0.48\textwidth]{RealCS+GR2.eps} 
\includegraphics[width=0.48\textwidth]{ImagCS+GR2.eps} 
\caption{Real (left panel) and imaginary (right panel) parts of free-space 
and in-medium scattering amplitudes at the $\eta N$ subthreshold region in 
model NLO30$_{\eta}$ of CS \cite{CS13} and as used in the $\eta$-nuclear 
bound-state calculation of GR \cite{GR02}. Free-space amplitudes are marked 
by 0 within brackets, nuclear-matter amplitudes accounting for Pauli-blocking 
and self-energies are marked by $\rho_0$ within brackets.} 
\label{fig:CS+GR} 
\end{center} 
\end{figure} 

The only other available calculations of $\eta$-nuclear bound states in nuclei 
using a coupled-channel in-medium model that accounts for Pauli blocking 
and for self-energies are due to Garc\'{i}a-Recio (GR) et al. \cite{GR02}. 
These calculations are based on a coupled channels approach developed in 
Ref.~\cite{IOV02} and its in-medium implementation in Ref.~\cite{IO02}. 
The GR underlying free-space and in-medium $\eta N$ subthreshold 
scattering amplitudes with Pauli blocking and self-energies are shown in 
Fig.~\ref{fig:CS+GR}, compared to those of CS that are used in the present 
work.   
%The free-space $\eta N$ scattering length derived in this model is 
%$a_{\eta N}=0.264+{\rm i}0.245$~fm \cite{IOV02}. 
We note that the GR in-medium real part is about $70\%$ higher at $\rho_0$ 
than the free-space real part \cite{IO02}, contrary to what is found by us 
for medium modifications based on model NLO30$_{\eta}$ of CS \cite{CS13}. 
We have no explanation for this disagreement between the two approaches 
except to recall that they differ appreciably in some of the self-energies 
input, notably for pions and $\Sigma$ hyperons. A partial resolution is 
offered by reversing Re\;$V_0^{\Sigma}$ in the CS in-medium evaluation from 
the value +30 MeV listed in Table~\ref{tab:depth} to the unrealistic value 
$-$30 MeV assumed by Inoue and Oset \cite{IO02}. This increases appreciably 
Re\;$F^{\rm CS}_{\eta N}(\sqrt{s},\rho_0)$ so that it almost reaches the 
level of its free-space counterpart in the subthreshold region. 
Im\;$F^{\rm CS}_{\eta N}(\sqrt{s},\rho_0)$ too increases appreciably, 
exceeding substantially its free-space counterpart and reaching the level of 
Im\;$F^{\rm GR}_{\eta N}(\sqrt{s},\rho_0)$ as constructed in Ref.~\cite{IO02}. 
(The case for {\it repulsive} Re\;$V_0^{\Sigma}$ is reviewed in 
Ref.~\cite{FG07}.) We note furthermore that these differences in going from 
free-space amplitudes to in-medium amplitudes between GR and CS appear already 
at the level of imposing Pauli blocking without recourse to SEs. In the GR 
calculations Pauli blocking {\it increases} the $\eta N$ free-space attraction 
at threshold according to \cite{IO02}, whereas in our CS-based model 
calculations it {\it decreases} this attraction in remarkable agreement with 
applying the multiple-scattering modification of Eqs.~(\ref{eq:WRW}) and 
(\ref{eq:WRWFG}). 

\begin{figure}[htb]
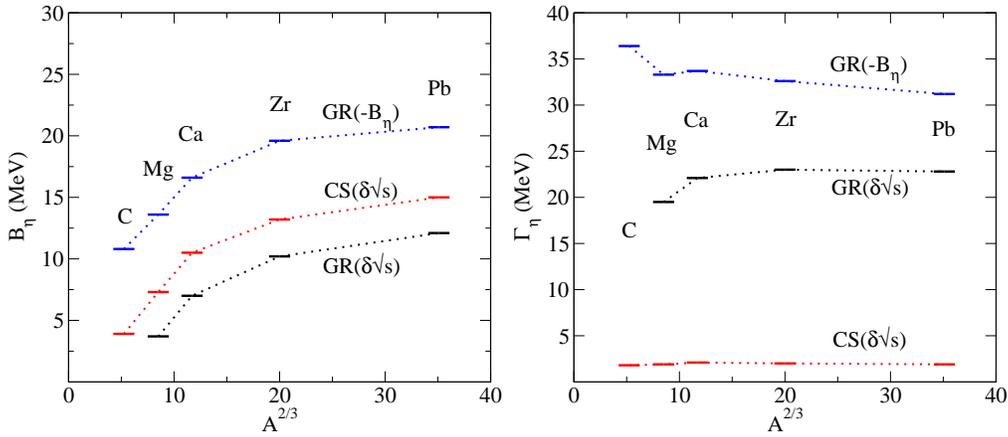
 
\begin{center} 
\includegraphics[width=0.48\textwidth]{beta-ales-oset.eps} 
\includegraphics[width=0.48\textwidth]{gamma-ales-oset.eps} 
\caption{Binding energies (left panel) and widths (right panel) of 
$1s_{\eta}$ bound states in nuclei calculated using the GR in-medium 
amplitudes \cite{GR02} with different procedures for handling 
self-consistently the subthreshold energy shift: $\delta\sqrt{s}$ stands 
for Eq.~(\ref{eq:sqrts}) and $-B_{\eta}$ stands for the procedure applied 
originally by GR. Shown also are results using NLO30$_{\eta}$ in-medium 
amplitudes marked CS \cite{CS13} with the present $\delta\sqrt{s}$ procedure 
Eq.~(\ref{eq:sqrts}).} 
\label{fig:ales-oset} 
\end{center} 
\end{figure} 

In spite of the differences in the underlying models, it is instructive to 
apply our self-consistency scheme of calculating $\eta$-nuclear bound states, 
based on $\delta\sqrt{s}$ of Eq.~(\ref{eq:sqrts}), to the GR in-medium 
energy-dependent and density-dependent $\eta N$ interaction, and to 
compare the results with those obtained by GR using a density-independent 
$\delta\sqrt{s}=-B_{\eta}$ self-consistency requirement. 
This comparison is made in Fig.~\ref{fig:ales-oset} where 
the in-medium NLO30$_{\eta}$ model results are included (denoted CS) using 
Eq.~(\ref{eq:sqrts}) for subthreshold energy values (marked $\delta\sqrt{s}$ 
in the figure). The left and right panels exhibit $1s_{\eta}$-nuclear binding 
energies and widths, respectively. All calculations include self-energies and 
coupled-channels evaluation of Pauli blocking. 

Comparing binding-energy and width results obtained by applying different 
self-consistency procedures, as presented in Fig.~\ref{fig:ales-oset}, 
one sees that our $\delta\sqrt{s}$ Eq.~(\ref{eq:sqrts}) procedure reduces 
considerably the GR binding energies and widths with respect to the original 
calculations that used a $\delta\sqrt{s}=-B_{\eta}$ procedure. However, even 
the reduced GR widths are still quite high, 20 MeV and over, suggesting that 
$\eta$-nuclear states will be prohibitively difficult to resolve if the GR 
model is the physically correct one. 

Considering the CS results one notes the remarkable smallness of the 
calculated widths shown on the right panel of Fig.~\ref{fig:ales-oset}, 
with values about 2 MeV. These very small widths do not include contributions 
from two-nucleon processes which are estimated to add a few MeV. We therefore 
anticipate that $1s_{\eta}$ and, wherever bound, also $1p_{\eta}$ nuclear 
states could in principle be observed if model NLO30$_{\eta}$ turns out to 
prove a realistic model. 

\begin{figure}[htb]
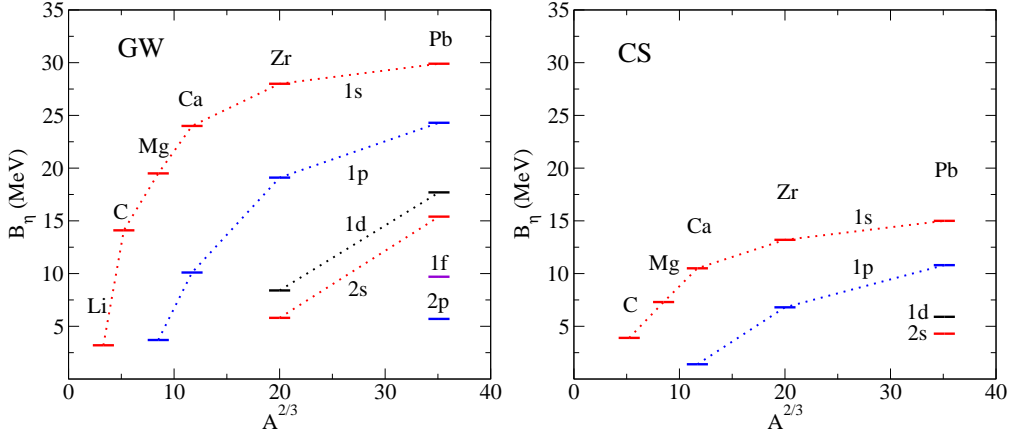
 
\begin{center} 
\includegraphics[width=0.48\textwidth]{beta2.eps} 
\includegraphics[width=0.48\textwidth]{beta-ales+se2.eps} 
\caption{Spectra of $\eta$-nuclear single-particle bound states across the 
periodic table, calculated self-consistently using in-medium models of the 
$\eta N$ subthreshold scattering amplitude, are shown in the left panel for 
the GW model \cite{GW05} and in the right panel for the NLO30$_{\eta}$ model 
of CS \cite{CS13}. Pauli blocking is included for both in-medium models, 
whereas hadron self-energies are accounted for only in the CS-based 
calculations.} 
\label{fig:GW-CS} 
\end{center} 
\end{figure} 

Finally, in Fig.~\ref{fig:GW-CS} we compare $\eta$-nuclear single-particle 
spectra across the periodic table evaluated self-consistently using two 
in-medium models, GW \cite{GW05} (left panel) and NLO30$_{\eta}$ due to 
CS \cite{CS13} (right panel). The free-space $\eta N$ amplitudes for 
these models may be viewed in Fig.~\ref{fig:aEtaN1.eps}. These dynamical 
calculations include Pauli blocking, using Eqs.~(\ref{eq:WRW}) and 
(\ref{eq:WRWFG}) for GW, and the coupled-channel approach discussed in 
Sect.~\ref{sec:etaN} for CS. The latter model also incorporates in-medium 
hadron self-energies, resulting in 2--3 MeV lower binding energies 
(see Fig.~5). The widths calculated in both models are remarkably small, 
as shown in Fig.~\ref{fig:ales-oset} for CS. For these two in-medium models 
$\eta$-nuclear single-particle bound states stand a chance of being observed, 
provided a suitable production/formation reaction is found. Other models 
studied by us produce either prohibitively large widths or are too weak to 
generate $\eta$-nuclear bound states over a substantial range of the periodic 
table.

\section{Summary and Outlook} 
\label{sec:sum} 

We have extended in the present work the self-consistent calculations 
of $\eta$-nuclear bound states reported recently in Ref.~\cite{FGM13} 
by using $\eta N$ scattering amplitudes that follow from the chirally-inspired 
meson-baryon coupled-channel model NLO30$_{\eta}$ \cite{CS13}. Pauli blocking 
and hadron self-energies are accounted for in the construction of in-medium 
amplitudes that serve as self-energy input to the $\eta$-nuclear KG equation 
for bound states. These amplitudes are both energy- and density-dependent, 
decreasing as one goes deeper into subthreshold for fixed density, as shown 
in Fig.~\ref{fig:aEtaNrho}. The in-medium subthreshold amplitudes encountered 
in $\eta$-nuclear bound-state calculations are substantially {\it weaker} 
both in their real part as well as in their imaginary part than the $\eta N$ 
scattering length. We have displayed in Fig.~\ref{fig:srhocs} the correlation 
in models CS \cite{CS13} and GW \cite{GW05} between the subthreshold energy 
downward shift $\delta\sqrt{s}$ and the nuclear density $\rho$ implied 
by satisfying Eq.~(\ref{eq:sqrts}). The resulting energy shifts of 
$-$(55$\pm$10)~MeV for central nuclear densities surpass considerably the 
shift $\delta\sqrt{s}=-B_{\eta}$ used in other works \cite{GR02,JNH02}. 
This is reflected in our calculated bound-state energies and widths 
which are smaller than those calculated in comparable models using 
$\delta\sqrt{s}=-B_{\eta}$, as shown in Fig.~\ref{fig:ales-oset} here. 
It is safe to conclude that irrespective of the underlying two-body $\eta N$ 
interaction model, a self-consistent treatment which couples together the 
energy dependence of the in-medium $\eta N$ scattering amplitude below 
threshold and the density dependence of the $\eta$-nuclear self-energy 
is mandatory in any future calculation of $\eta$-nuclear bound states. 

The small $\eta$-nuclear widths of 2--3 MeV calculated in the CS in-medium 
model, and also in the GW in-medium model, might encourage further 
experimental activity seeking to produce and identify $\eta$-nuclear bound 
states. These small widths, however, are model dependent, as evidenced by 
the substantially larger widths calculated in other models as displayed 
in Fig.~\ref{fig:ales-oset}. Additional width contributions disregarded 
in our in-medium model are due to two-pion production $\eta N\to\pi\pi N$ 
and two-nucleon absorption $\eta NN\to NN$. These contributions 
are estimated to add a few MeV to $\eta$-nuclear widths evaluated in 
meson-baryon coupled-channels approaches, so we feel it is safe to 
assume that the {\it total} $\eta$-nuclear widths in model NLO30$_{\eta}$ 
do not exceed 5--10 MeV. To appreciate the smallness of these 
estimated widths, we recall the semiclassical estimate 
$\Gamma^{\rm QF}_{\eta}\approx v_{\eta}\sigma^{\rm abs}_{\eta N}\rho_0$, 
where $v_{\eta}=p_{\eta}/E_{\eta}$ and $\sigma^{\rm abs}_{\eta N}=(30\pm 
2.5_{\rm stat}\pm 6_{\rm syst})$ mb is the $\eta$-meson absorption cross 
section in nuclear matter as determined from near threshold photoproduction 
of quasi-free (QF) $\eta$ mesons on complex nuclei at MAMI \cite{MAMI96}. 
Using the lowest $\eta$-meson kinetic energy, $T_{\eta}\approx 25$ MeV, 
at which this determination of $\sigma^{\rm abs}_{\eta N}$ appears stable, 
one gets $\Gamma^{\rm QF}_{\eta}\approx (29\pm 6)$~MeV. 
Of course, given the rapid fall-off of Im\;$F_{\eta N}(\sqrt{s})$ as 
$\sqrt{s}\to\sqrt{s_{\eta N}}$, this semiclassical QF estimate cannot be 
reliably extrapolated down to threshold and would certainly break down in 
the $\eta$-nuclear subthreshold region where the calculations presented 
here are performed. We recall, furthermore, that only a single claim of 
observing $\eta$-nuclear bound states has been made to date, in the reaction 
$p+{^{27}{\rm Al}}\to {^3{\rm He}}+{_{\eta}^{25}{\rm Mg}}\to {^3{\rm He}}+p+
\pi^-+X$ as reported recently by the COSY-GEM collaboration \cite{GEM09}. 
The width extracted for the claimed peak is $\Gamma({_{\eta}^{25}{\rm Mg}})=
(10\pm 3)$~MeV. For updated overview of past, present and future in 
$\eta$-nuclear experiments and theory, we refer to the recent symposium 
on Mesic Nuclei at Cracow, Sept. 2014 \cite{moskal13}. 

The subthreshold behavior of $s$-wave meson-baryon scattering amplitudes 
and its consequences for meson-nuclear bound states has been dealt by us 
extensively for $K^-$ mesons, see Refs.~\cite{Gal13,Gal14} for recent reviews. 
Another meson-baryon system of interest is $\eta'(958)N$. The QCD connection 
between $\eta$- and $\eta'$-nuclear bound states has been highlighted recently 
by Bass and Thomas, emphasizing predictions of the QMC model \cite{QMC13}. 
Experimentally, a value of $\Gamma_{\eta'}(\rho_0)$=$(20\pm 5)$~MeV for the 
in-medium $\eta'$-meson width was derived from measured transparency ratios 
in $\eta'$ photoproduction on nuclei \cite{Nanova12}. Very recently, a value 
of $V_{\eta'}(\rho_0)$=$-(37 \pm 10_{\rm stat} \pm 10_{\rm syst})$ MeV for 
the real part of the $\eta'$-nuclear potential depth has been determined by 
measuring the $\eta'$-meson excitation function and momentum distribution 
in photoproduction on $^{12}$C \cite{Nanova13}. It is worth noting however 
that the high-momentum $\eta'$ mesons produced in these photoproduction 
experiments, with $p_{\eta'}\sim 1$~GeV/c, are kinematically far away from 
the low-momentum range expected for meson-nuclear bound-state systems. 
Furthermore, the rather strong in-medium $\eta' N$ attraction and 
absorption derived from these experiments is at odds with the value 
$|a_{\eta' N}|\approx 0.1$~fm derived from the near-threshold $pp\to pp\eta'$ 
reaction \cite{Moskal00}. Dedicated experiments are planned to 
search for $\eta'$-nuclear bound states in $(\pi^+,p)$ or $(p,d)$ 
reactions \cite{Nagahiro13,Itahashi13}. Yet, the issue of energy dependence 
of the $\eta' N$ scattering amplitude and its relevance for the calculation 
of $\eta'$-nuclear bound states has not been considered, and in view of the 
results reported here for $\eta$-nuclear bound states it is of considerable 
interest to follow in near-future work.

\section*{Acknowledgements} 
This work was supported by the GACR Grant No. 203/12/2126, as well as 
by the EU initiative FP7, HadronPhysics3, under the SPHERE and LEANNIS 
cooperation programs. J.M. acknowledges financial support within the 
agreement on scientific collaboration between the Academy of Sciences 
of the Czech Republic and the Israel Academy of Sciences and Humanities. 
Both A.C. and J.M. acknowledge the hospitality extended to them at the 
Racah Institute of Physics, The Hebrew University of Jerusalem, during 
a collaboration visit in April 2013.

\end{document}